\documentclass[10pt,twocolumn]{IEEEtran}
\usepackage{graphicx}           % Important!!
\usepackage{amsmath,amsfonts}
\usepackage{multirow}

\begin{document}

\title{Toward Cloud-based Vehicular Networks with Efficient Resource Management}

\author{Rong Yu, \emph{Member, IEEE}, Yan Zhang, \emph{Senior Member,
IEEE},  Stein Gjessing,~\IEEEmembership{Senior Member, IEEE,} \\Wenlong Xia,
\emph{Student Member, IEEE}, Kun Yang, ~\IEEEmembership{Senior Member, IEEE}

\thanks{Rong Yu and Wenlong Xia are with Guangdong University of Technology, China.
Email: yurong@ieee.org, wlxia@ieee.org}
\thanks{Yan Zhang and Stein Gjessing are with Simula Research Laboratory, Norway;
    and University of Oslo, Norway.
    Email: yanzhang@ieee.org, steing@ifi.uio.no}
\thanks{Kun Yang is with University of Essex, UK.
    Email: kunyang@essex.ac.uk}
}

\maketitle

\begin{abstract}
In the era of Internet of Things, all components in intelligent
transportation systems will be connected to improve transport safety, relieve
traffic congestion, reduce air pollution and enhance the comfort of driving.
The vision of \textit{all vehicles connected} poses a significant challenge
to the collection and storage of large amounts of traffic-related data. In
this article, we propose to integrate cloud computing into vehicular networks
such that the vehicles can share computation resources, storage resources and
bandwidth resources. The proposed architecture includes a vehicular cloud, a
roadside cloud, and a central cloud. Then, we study cloud resource allocation
and virtual machine migration for effective resource management in this
cloud-based vehicular network. A game-theoretical approach is presented to
optimally allocate cloud resources. Virtual machine migration due to vehicle
mobility is solved based on a resource reservation scheme.
\end{abstract}

\begin{keywords}
cloud computing, mobile cloud, vehicular networks, Internet of vehicles,
resource management, cloudlet, virtual machine migration.
\end{keywords}

\section{Introduction}

Vehicular networks are in the progress of merging with the Internet to
constitute a fundamental information platform which is an indispensable part
of Intelligent Transport System (ITS) \cite{IoV}. This will eventually evolve into all vehicles connected in the era of Internet of Things (IoT)
\cite{ITU-Report}. By supporting traffic-related data gathering and
processing, vehicular networks is able to notably improve transport safety,
relieve traffic congestion, reduce air pollution, and enhance driving
comfortability \cite{Chenjm11}. It has been reported that, in Western Europe, 25\% of the
deaths due to car accidents could be reduced by deploying safety warning
systems at the highway intersections \cite{WHO}. Another example is that
real-time traffic information could be collected and transmitted to data
center for processing, and in return, information could be broadcasted to the drivers for route planning. City traffic congestion is alleviated and
traveling time is reduced, leading to greener cities.

A variety of information technologies have been developed for intelligent vehicles, roads, and traffic infrastructures such that all vehicles are connected. Smart sensors and actuators are deployed in vehicles and roadside infrastructures for data acquisition and decision. Advanced communication technologies are used to interconnect vehicles and roadside infrastructures, and eventually access to Internet. For instance, Dedicated Short Range Communications (DSRC) is specifically designed for Vehicle-to-Vehicle (V2V) and Vehicle-to-Roadside (V2R) communications. The IEEE 802.11p, called Wireless Access in Vehicular Environments (WAVE) \cite{WAVE}, is currently a popular standard for DSRC. Besides, the Long-Term Evolution (LTE), LTE-Advanced and Cognitive Radio (CR) \cite{Yu12}\cite{Liu10} are all fairly competitive technologies for vehicular networking \cite{SongTVT}\cite{SongJSAC}.

Despite of the well-developed information technologies, there is a significant challenge that hinders the rapid development
of vehicular networks. Vehicles are normally constrained by resources,
including computation, storage, and radio spectrum bandwidth. Due to the
requirements of small-size and low-cost hardware systems, a single vehicle
has limited computation and storage resources, which may result in low data
processing capability. On the other hand, many emerging applications demands
complex computation and large storage, including in-vehicle multimedia
entertainment, vehicular social networking, and location-based services. It
becomes increasingly difficult for an individual vehicle to efficiently
support these applications. A very promising solution is to share the
computation and storage resources among all vehicles or physically nearby
vehicles. This motivates us to study the new paradigm of cloud-based vehicular networks.

\begin{figure*}[t]
\centering
  \includegraphics[width=0.95\textwidth]{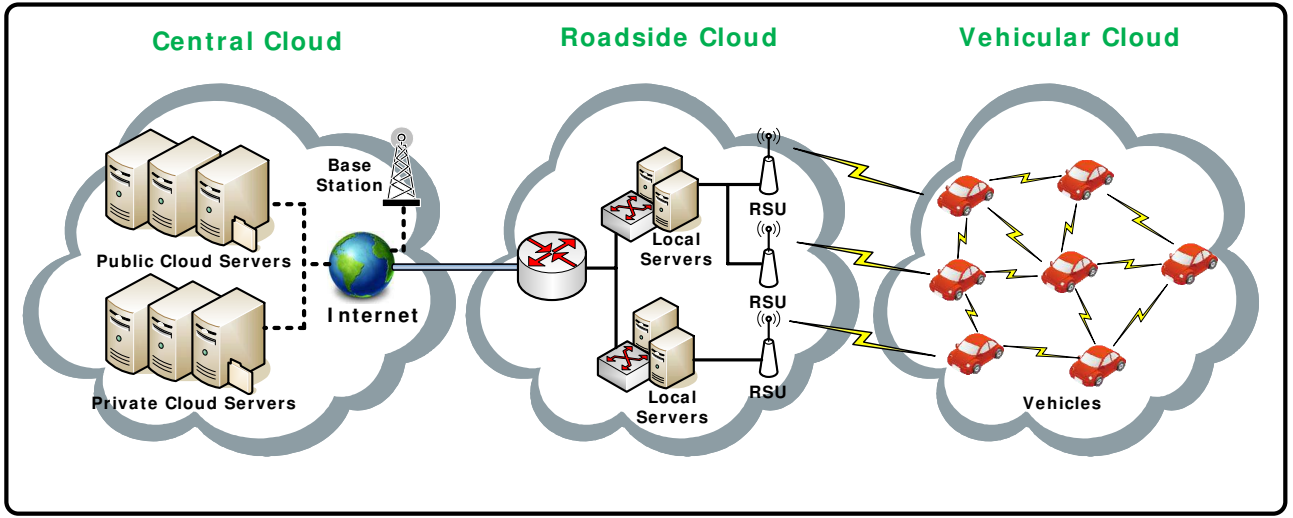}\\
  \caption{Proposed cloud-based vehicular networks architecture}
  \label{fig:home area network architecture}
\end{figure*}

Recently, a few research works are reported to study the combination of cloud computing and vehicular networks. In \cite{Olariu2011}, the concept of Autonomous Vehicular Clouds (AVC) is proposed to exploit the under-utilized resources in vehicular ad hoc networks (VANETs). A Platform as a Service (PaaS) model is designed in \cite{Paas_VANET} to support cloud services for mobile vehicles. The work in \cite{VANET_Cloud} proposes architectures of Vehicular Clouds (VC), Vehicles using Clouds (VuC) and Hybrid Clouds (HC). Vehicles play as cloud service providers and clients respectively in VC and VuC, and as hybrids in HC.

In this article, we propose a hierarchical cloud architecture for vehicular networks.
Our work is different from previous researches in three main aspects.
First, we aim to create a pervasive cloud environment for mobile vehicles by integrating redundant physical resources in ITS infrastructures, including data center, roadside units and vehicles. The aggregation of these sporadic physical resources potentially compose massive and powerful cloud resources for vehicles.
Second, we propose a three-layered architecture to organize the cloud resources.  The layered structure allows vehicles to select their cloud services resiliently. Central clouds have sufficient cloud resources but large end-to-end communications delay. On the contrary, roadside cloud and vehicular cloud have limited cloud resources but satisfying communications quality.
Third, we emphasize the efficiency, continuity and reliability of cloud services for mobile vehicles. As a consequence, efficient cloud resources management strategies are elaborately proposed. Countermeasures to deal with vehicle mobility are devised.

The remainder of the article is organized as follows. Section II illustrates the proposed architecture that includes vehicular cloud, roadside cloud, and central cloud. Cloud deployment strategies are discussed for these three layers. Section III envisions several promising applications for different resources sharing in cloud-based vehicular networks. In Section IV, we focus on cloud resource allocation problems and a game-theoretical approach is presented to optimally allocate cloud resources. In Section V, we study virtual machine migration due to vehicle mobility. Illustrative results indicate resource allocation optimization and virtual machine migration performance. The conclusion is presented in Section VI.

\section{Proposed Cloud-based Vehicular Networks Architecture}

Fig.1 shows the proposed cloud architecture for vehicular networks. It is a hierarchical architecture that consists of three interacting layers:  vehicular cloud, roadside cloud, and central cloud. Vehicles are mobile nodes that exploit cloud resources and services.
\begin{itemize}
    \item \emph{Vehicular Cloud}: a local cloud established among a group of cooperative vehicles. An inter-vehicle network, i.e., a vehicular ad hoc network (VANET), is formed by V2V communications. The vehicles in a group are viewed as mobile cloud sites and they cooperatively create a vehicular cloud.
    \item \emph{Roadside Cloud}: a local cloud established among a set of adjacent roadside units. In a roadside cloud, there are dedicated local cloud servers attached to Roadside Units (RSUs). A vehicle will access a roadside cloud by V2R communications.
    \item \emph{Central Cloud}: a cloud established among a group of dedicated servers  in the Internet. A vehicle will access a central cloud by V2R or cellular communications.
\end{itemize}

This architecture has several essential advantages. First, the architecture
fully utilizes the physical resources in an entire network. From vehicles to
roadside infrastructures and data center, the computation and storage
resources are all merged into the cloud. All clouds are accessible to all
vehicles. Second, the hierarchical nature of the architecture allows vehicles using different communication technologies to access to different layers of clouds accordingly. Hence, the architecture is flexible and
compatible with heterogeneous wireless communication technologies, e.g., DSRC, LTE/LTE-Advanced and CR technologies. Third, the
vehicular clouds and the roadside clouds are small-scale localized clouds.
Such distributed clouds can be rapidly deployed and provide services quickly.

\subsection{Vehicular Cloud}

In a vehicular cloud, a group of vehicles share their computation resources,
storage resources, and spectrum resources. Each vehicle can access the cloud
and utilize services for its own purpose. Through the cooperation in the
group, the physical resources of vehicles are dynamically scheduled on
demand. The overall resource utilization is significantly enhanced. Compared
to an individual vehicle, a vehicular cloud has much more resources.

Due to vehicle mobility, vehicular cloud implementation is very different
from a cloud in a traditional computer network. We propose two customization
strategies for vehicular clouds: Generalized Vehicular Cloud Customization
(GVCC) and Specified Vehicular Cloud Customization (SVCC).

In GVCC, a \emph{cloud controller} is introduced in a vehicular cloud. A cloud
controller is responsible for the creation, maintenance, and deletion of a
vehicular cloud. All vehicles will virtualize their physical resources and
register the virtual resources in the cloud controller. All virtual resources
of the vehicular cloud are scheduled by the cloud controller. If a vehicle
needs some resources of the vehicular cloud, it should apply to the cloud
controller. In contrast to GVCC, SVCC has no cloud controller. A vehicle will
specify some vehicles as candidate cloud sites, and directly apply for
resources from these vehicles. If the application is approved, the
corresponding vehicles become cloud sites, which will customize virtual
machines (VMs) according to the vehicle demand.

These two strategies, GVCC and SVCC, are quite different. With respect to
resource management, GVCC is similar to a conventional cloud deployment
strategy in which cloud resources are scheduled by a controller. A vehicle is
not aware of the cloud sites where the VMs are built up. The cloud controller
should maintain the cloud resources. During a cloud service, if a cloud site
is not available due to vehicle mobility, the controller should schedule a
new site to replace it. In SVCC, since there is no cloud controller, a
vehicle has to select other vehicles as cloud sites and maintain the cloud
resources itself. In terms of resource utilization, GVCC is able to globally
schedule and allocate all resources of a vehicular cloud. GVCC has higher
resource utilization than SVCC. However, the operation of the cloud
controller will need extra computation. Therefore, SVCC may be more efficient
than GVCC in terms of lower system overhead.

\begin{figure*}[t]
\centering
  \includegraphics[width=1.0\textwidth]{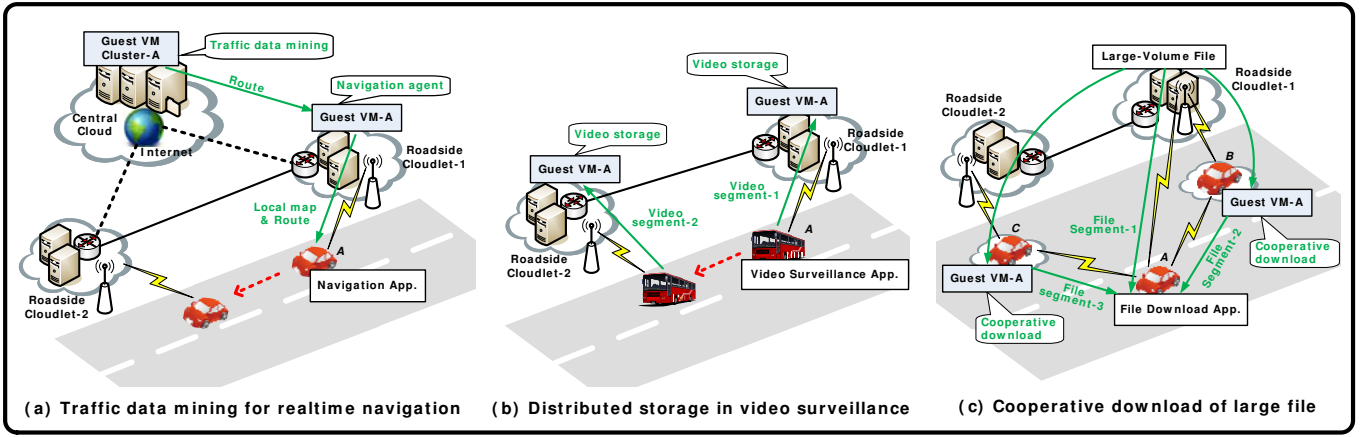}\\
  \caption{Applications of cloud-based vehicular networks}
\end{figure*}

\subsection{Roadside Cloud}

A roadside cloud is composed of two main parts: dedicated local servers and
roadside units. The dedicated local servers virtualize physical resources and
act as a potential cloud site. RSUs provide radio interfaces for vehicles to
access the cloud. A roadside cloud is accessible only by the nearby vehicles,
i.e., those locate within the radio coverage area of the cloud site's RSU.
This fact helps us recall the concept of a \textit{cloudlet}. A cloudlet is a
trusted, resource-rich computer or cluster of computers that is connected to
the Internet and is available for use by nearby mobile devices
\cite{cloudlet}. In this article, we propose the concept of a roadside
cloudlet. A roadside cloudlet refers to a small-scale roadside cloud site
that offers cloud services to bypassing vehicles. A vehicle can select a
nearby roadside cloudlet and customize a transient cloud for use. Here, we
call the customized cloud a \emph{transient cloud} because the cloud can only
serve the vehicle for a while. After the vehicle moves out of the radio range
of the current serving RSU, the cloud will be deleted and the vehicle will
customize a new cloud from the next roadside cloudlet in its moving
direction.

When a vehicle customizes a transient cloud from a roadside cloudlet, it is
offered by virtual resources in terms of virtual machine (VM). This VM
consists of two interacting components: the VM-base in the roadside cloudlet
and the VM-overlay in the vehicle. A VM-base is a resource template recording
the basic structure of a VM, while a VM-overlay mainly contains the specific
resource requirements of the customized VM. Before a cloud service starts,
the vehicle will send the VM-overlay to the roadside cloudlet. After
combining the VM-overlay with the VM-base, the roadside cloudlet completes
the customization of a dedicated VM. During a cloud service, as the vehicle
moves along the roadside, it will switch between different RSUs. For the
continuity of cloud service, the customized VM should be synchronously
transferred between the respective roadside cloudlets. This process is
referred to as VM migration. VM migration scenarios will be further
elaborated in Section V.

\subsection{Central Cloud}

Compared to a vehicular cloud and a roadside cloud, a central cloud has much
more resources. The central cloud can be driven by either dedicated servers
in vehicular networks data center or servers in the Internet. A central cloud
is mainly used for complicated computation, massive data storage, and global
decision. There already exist mature open source or commercial software
platforms that could be employed for the deployment of a central cloud.
Openstack is an open source cloud platform using Infrastructure as a Service
(IaaS) model. Other potential commercial platforms are Amazon Web Services,
Microsoft Azure and Google App Engine.

\section{Promising Applications of Cloud-based vehicular networks}

With powerful cloud computing, cloud-based vehicular networks can support
many unprecedented applications. In this section, we illustrate potential
applications and explain the exploitation of a vehicular cloud, a roadside
cloud, and a central cloud to facilitate new applications.

\subsection{Realtime Navigation with Computation Resources Sharing}

In a realtime navigation application, the computation resources in the
central cloud is utilized for traffic data mining. Vehicles may offer
services that use resources that are outside their own computing ability.
Different from traditional navigation that can only provide static geographic
maps, realtime navigation is able to offer dynamic three-dimensional maps and
adaptively optimize routes based on traffic data mining.

In Fig.~2(a), vehicle $A$ is using realtime navigation during its traveling.
It will first request cloud service from the central cloud and the roadside
cloud. Then, a VM cluster and a VM are established in the central cloud and
the roadside cloud, respectively. The VM cluster-A in the central cloud is in
charge of traffic data mining and will suggest several routes based on the
current traffic conditions. Once a route is selected by $A$, realtime
navigation starts. VM-A in the roadside cloud acts as an agent to push
messages to vehicle $A$, updating the driver with traffic conditions on the
road. As vehicle $A$ moves on, VM-A will migrate to different roadside cloud
sites. During the entire travel, VM cluster-A in the central cloud will keep
updating the route information based on realtime traffic condition. Once
there is an unexpected event, e.g., traffic congestion, VM cluster-A will
report the situation timely and compute a new route.

\begin{table*}[t]
\renewcommand{\arraystretch}{1.3}
\caption{Applications of Cloud-based vehicular networks}\centering
\tabcolsep=6pt
\begin{tabular}{c|c|c|c|c|c|c}
\hline
\multicolumn{1}{c|}{\multirow {2}{*}{Potential Applications}}& \multicolumn{3}{|c|}{Relevant Cloud Assistance}&\multicolumn{3}{|c}{Resource Sharing}\\
\cline{2-7} \multicolumn{1}{c|}{}&
\multicolumn{1}{|c|}{Central Cloud} &\multicolumn{1}{|c|}{Roadside Cloud} &\multicolumn{1}{|c|}{Vehicular Cloud}&\multicolumn{1}{|c|}{Computation} &\multicolumn{1}{|c|}{Storage} &\multicolumn{1}{|c}{Bandwidth}\\
\hline \multicolumn{1}{l|}{Realtime Traffic Condition Analysis and
Broadcast}&
\multicolumn{1}{|c|}{$\surd$}&\multicolumn{1}{|c|}{$\surd$}&\multicolumn{1}{|c|}{}& \multicolumn{1}{|c|}{$\surd$}&\multicolumn{1}{|c|}{$\surd$}&\multicolumn{1}{|c}{} \\
\hline \multicolumn{1}{l|}{Realtime Car Navigation}&
\multicolumn{1}{|c|}{$\surd$}&\multicolumn{1}{|c|}{$\surd$}&&
\multicolumn{1}{|c|}{$\surd$}&\multicolumn{1}{|c|}{}&\multicolumn{1}{|c}{} \\
\hline \multicolumn{1}{l|}{Video Surveillance}&
\multicolumn{1}{|c|}{}&\multicolumn{1}{|c|}{$\surd$}&\multicolumn{1}{|c|}{}&
\multicolumn{1}{|c|}{}&\multicolumn{1}{|c|}{$\surd$}&\multicolumn{1}{|c}{} \\
\hline \multicolumn{1}{l|}{LBS Commercial Advertisement}&
\multicolumn{1}{|c|}{}&\multicolumn{1}{|c|}{$\surd$}&\multicolumn{1}{|c|}{$\surd$}&
\multicolumn{1}{|c|}{}&\multicolumn{1}{|c|}{$\surd$}&\multicolumn{1}{|c}{$\surd$}  \\
\hline \multicolumn{1}{l|}{Vehicular Mobile Social Networking}&
\multicolumn{1}{|c|}{$\surd$}&\multicolumn{1}{|c|}{}&\multicolumn{1}{|c|}{$\surd$}&
\multicolumn{1}{|c|}{$\surd$}&\multicolumn{1}{|c|}{}&\multicolumn{1}{|c}{$\surd$} \\
\hline \multicolumn{1}{l|}{In-Vehicle Multimedia Entertainment}&
\multicolumn{1}{|c|}{}&\multicolumn{1}{|c|}{$\surd$}&$\surd$&
\multicolumn{1}{|c|}{$\surd$}&\multicolumn{1}{|c|}{$\surd$}&\multicolumn{1}{|c}{$\surd$} \\
\hline \multicolumn{1}{l|}{Inter-Vehicle Video and Audio Communications}&
\multicolumn{1}{|c|}{}&\multicolumn{1}{|c|}{}&$\surd$&
\multicolumn{1}{|c|}{}&\multicolumn{1}{|c|}{}&\multicolumn{1}{|c}{$\surd$} \\
\hline \multicolumn{1}{l|}{Remote Vehicle Diagnosis}&
\multicolumn{1}{|c|}{$\surd$}&\multicolumn{1}{|c|}{$\surd$}&\multicolumn{1}{|c|}{}&
\multicolumn{1}{|c|}{$\surd$}&\multicolumn{1}{|c|}{}&\multicolumn{1}{|c}{} \\
\hline
\end{tabular}
\end{table*}

\subsection{Video Surveillance With Storage Resource Sharing}

Video surveillance is an important application that utilizes shared storage
resources. Currently, many buses in a city have installed High-Definition
(HD) camera systems to monitor in-bus conditions. A very large-volume
hard-drive is needed to store video content for a couple of days. This video
storage scheme has several disadvantages. First, to save HD video content for
days, the hard-drive should have very large storage, which leads to high cost
and big size. Second, video content can only be checked in an off-line
manner, the department of transportation is not able to make timely and
proper decisions immediately after an accident. In cloud-based vehicular
networks, a new distributed storage paradigm can address this problem. The
storage capability of in-bus video camera systems is significantly extended.

In Fig.~2(b), bus $A$ exploits the roadside cloud to facilitate storage of
in-bus video surveillance content. Specifically, the bus applies for cloud
services and receives a VM in the roadside cloud. The video content is
uploaded to the guest VM-A in the roadside cloudlet-1 in a realtime manner.
When the bus moves along the road and is located in the coverage area of the
roadside cloudlet-2, VM-A will be migrated accordingly. As a result, the
video content is divided into several segments and separately stored in
different roadside cloudlets along the road. The video segments in the
roadside cloudlets will be transmitted to a data center on demand. When an
accident is reported, department of transportation can request roadside
cloudlets to send back video to the data center.

\subsection{Cooperative Download/Upload with Bandwidth Sharing}

Cooperative downloading and uploading services are interesting applications
that share bandwidth resources. Many new applications involves large-volume
data uploading or downloading. Typical examples include in-vehicle multimedia
entertainment, location-based rich-media advertisements, and big-size e-mail
services. Due to limited wireless bandwidth and vehicle movement, it is very
difficult to download an entire large file from a specific RSU. While the
vehicle drives by, there is not enough time to complete the download of large
amount of data. Here, we illustrate that the usage of a vehicular cloud will
make such applications feasible.

In Fig.~2(c), vehicle $A$ is going to download a large file from the roadside
infrastructure. The cooperative downloading has two phases. In the first
phase, vehicle $A$ observes neighboring vehicles $B$ and $C$ and then setups
a vehicular cloud for cooperative downloading. Then, a guest VM will be
constructed in both $B$ and $C$. The file downloading will be carried out by
the vehicular cloud that consists of vehicle $A$ and the two VMs on $B$ and
$C$. Since the file is downloaded by three vehicles in a parallel manner, the
total transmission rate becomes much faster. In this way, vehicle $A$ has
high possibility to finish downloading before vehicle $A$ moves out of the
range of the roadside infrastructure. In the second phase, the VMs in $B$ and
$C$ will further cooperatively transmit two separated segments of the file to
$A$. Since only V2V communications is involved, the second phase can be
performed without the roadside infrastructure. After that, $A$ will
reassemble the file segments into an entire file.

Table I summarizes potential applications in cloud-based vehicular networks.
We also show the relevant cloud resource sharing in each application.

\section{Game-theoretical Approach for Resource Allocation}

Vehicle and roadside clouds are both resource-intensive components. Resource
management is very crucial for these two types of clouds. Resources in
vehicle and roadside clouds are represented in form of VMs. In the literature, VM research has been mainly studied in computer networks. In the recent study \cite{MishraCOMMAG2012}, VM migration is considered for dynamic resource management in cloud
environments. In \cite{cloudspider}, VM replication and scheduling are
intelligently combined for VM migration across wide area network
environments. 
However, there are few studies on VM resource management in mobile cloud
environments. In \cite{cloudlet}, cloudlet is discussed and customized in the mobile computing environments.

In this section, we mainly focus on VM resource allocation in vehicular clouds
and roadside clouds. In a roadside cloud, there are multiple VMs since a
cloud site provides services to several vehicles simultaneously. In this
case, the resources in a cloud site should be appropriately allocated. VM
resource allocation should consider several aspects. i) Efficiency: VM resource allocation strategy should be efficient such that the limited resources are fully utilized. ii) Quality-of-Service (QoS): the allocated resources to a specific VM should be sufficient for the accomplishment of the VM's tasks to achieve its QoS requirements. iii) Fairness: VMs with the same workload should be offered statistically equal resources. Here, we formulate the competition among the VMs for cloud resources as a non-cooperative game.

\subsection{Game-theoretical Model}

Consider a roadside cloudlet with $N$ VMs, i.e., \emph{players} of the game. The VMs will
apply to the cloud site and compete for resources. These VMs are selfish in
the sense that they aim to obtain as much resources as possible for their own
usage. The cloud will allocate the total available resources to the VMs in
proportion to the number of requested resources.

Let $C$ and $M$ represent the total available computation and storage resources of the cloud site, respectively.
Let $c_i$ ($0<c_i\leq C$) and $m_i$ ($0<m_i\leq M$) denote the number of requested resources from the $i$-th VM in computation and storage, respectively.  Define $c_{-i}=\sum_{n=1,n\neq i}^N c_n$ and
$m_{-i}=\sum_{n=1,n\neq i}^N m_n$. The $i$-th VM will be allocated
computation and storage resources $\frac{c_i C}{c_i+c_{-i}}$ and $\frac{m_i
M}{m_i+m_{-i}}$, respectively. For the sake of fairness, the cloud site sets
up two Virtual Resource Counters (VRCs) for each VM. These two VRCs are used
to record the accumulative number of applied resources, one for computation
and the other for storage. When a VRC reaches its maximal value, the VM is
not allowed to apply for that type of resource. By using VRCs, the total
amount of allocated resources are equal for all VMs from a long-term
perspective. Let $\alpha_i$ and $\beta_i$ ($\alpha_i>0$, $\beta_i>0$),
respectively, denote the predefined resource weights that indicate the importance of
computation and storage resources in the workloads of the $i$-th VM, and let
$\lambda_i$ and $\gamma_i$ ($\lambda_i>0$, $\gamma_i>0$) denote the pricing
factors associated with applied computation and storage resources,
respectively, of the $i$-th VM. The utility function, or \emph{payoff}, for the $i$-th VM is given by
\begin{equation}\label{eqn:utility}
U(c_i,m_i) = \frac{\alpha_i c_i C}{c_i+c_{-i}}+\frac{\beta_i m_i
M}{m_i+m_{-i}}-\left(\lambda_i c_i+\gamma_i m_i\right).
\end{equation}

The proposed game-theoretical model is specially devised for mobile cloud applications in vehicular networks. In particular, the resource weights $\alpha_i$ and $\beta_i$ in the utility function make the game model adaptable to resources preference in different applications.
%These two weights are determined by the VM's demand on computation and storage resources.
The pricing factors $\lambda_i$ and $\gamma_i$ are set to prevent from resource waste imposed by excessive competition, and thus, potentially enhance resource utilization. These key parameters $\alpha_i$, $\beta_i$, $\lambda_i$ and $\gamma_i$ are elaborately selected regarding the mobile environment of the cloud-assisted vehicles.
For example, vehicles may have different quality of radio links to the cloud site. Their VMs should be provided with different $\alpha_i$, $\beta_i$, $\lambda_i$ and $\gamma_i$ according to the link quality. Typicaly, in a mobile multimedia application where scalable video coding (SVC) technique is involved, the VM is responsible for adaptive video decoding in the cloud site. The required VM resource mostly depends on the link quality. Because the link rate restricts the affordable quality of video stream, and consequently, determines the amount of VM resources for video processing.

\subsection{Nash Equilibrium}

In a non-cooperative game, a Nash equilibrium is a balanced state with a
strategy profile, from which no game player has any incentive to deviate. In
the proposed VM resource allocation game, by computing the second order
derivative of $U(c_i,m_i)$ with respect to $c_i$ and $m_i$ respectively, we
get $\frac{\partial^2 U}{{\partial c_i}^2} = -\frac{ 2\alpha_i
c_{-i}C}{(c_i+c_{-i})^3}<0$ and $\frac{\partial^2 U}{{\partial m_i}^2} =
-\frac{ 2\beta_i m_{-i}M}{(m_i+m_{-i})^3}<0$. This means that, $U(c_i,m_i)$
is a concave function with respect to $c_i$ or $m_i$. Therefore, the
existence of a Nash equilibrium is proven in the VM resource allocation game
model \cite{GameTheory}. Given the other VMs' applications, say, $c_{-i}$ and $m_{-i}$, we define $(c^*_i,m^*_i) \in \arg\max U(c_i,m_i)$ as the \emph{best response},t or called \emph{optimal strategy}, of the $i$-th VM in each iteration. We have
\begin{equation}\label{eqn:best-response}
\left\{
\begin{split}
& c^*_i = \min\left(C, \sqrt{\frac{\alpha_i c_{-i}C}{\lambda_i}}- c_{-i}\right),\\
& m^*_i = \min\left(M, \sqrt{\frac{\beta_i m_{-i} M}{\gamma_i}} -m_{-i}\right).
\end{split}\right.
\end{equation}
To prove the uniqueness of Nash equilibrium in the VM resource allocation
game, we can validate that the best response function is a standard
function, which has three features: positivity, monotonicity and scalability
\cite{GameTheory}. Follwing (\ref{eqn:best-response}), it is easy to prove that the sufficient conditions for the uniqueness of Nash equilibrium are $\forall i$, $\alpha_i\geq 4(N\!-\!1)\lambda_i$ and $\beta_i\geq 4(N\!-\!1)\gamma_i$.

\begin{figure}[t]
\centering
  \includegraphics[width=0.51\textwidth]{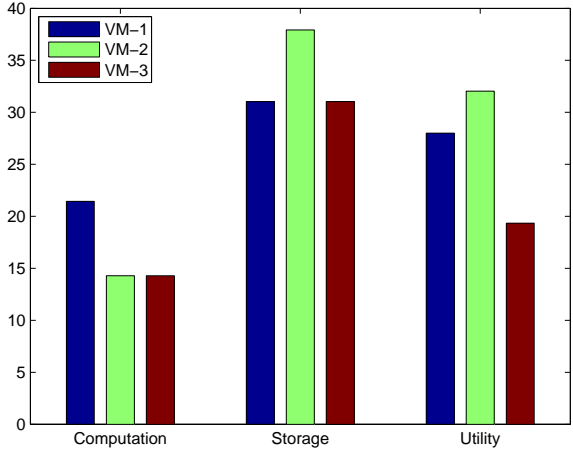}\\
  \caption{Resource allocation result in roadside cloud}
\end{figure}

Fig.3 shows a numerical example of resource allocation in our
game-theoretical model. In the example, there are three VMs in a roadside
cloud. The total available resources in computation and storage are set to
$50$ and $100$ units, respectively. The VMs have different resource demands.
VM-1 has the highest demand on computation while VM-2 has the highest demand
on storage.
%
%Thus, the weights in the game model are set as $\alpha_{\{1,2,3\}}=\{2,0.8,1.2\}$ and $\beta_{\{1,2,3\}}=\{1,2,1.5\}$.
%
%$\lambda_{\{1,2,3\}}=\{2,1,1\}$ and $\gamma_{\{1,2,3\}}=\{1,1.8,1.5\}$.
%
We randomly select initial values of the resource applications for the three VMs, say, $c_{\{1,2,3\}}=\{10,5,5\}$ and $m_{\{1,2,3\}}=\{5,15,10\}$. In the simulation, it is observed that the game iteration converges fast. The game reaches its Nash equilibrium after nearly $10$ rounds of iterations.
Results indicate that the resources are appropriately allocated
based on demand. In particular, VM-1, VM-2 and VM-3 are allocated $21.4,
14.3, 14.3$ units in computation, respectively. VM-1, VM-2 and VM-3 are
allocated $31.1, 37.8, 31.1$ units in storage, respectively.

\begin{figure*}[t]
\centering
  \includegraphics[width=1.0\textwidth]{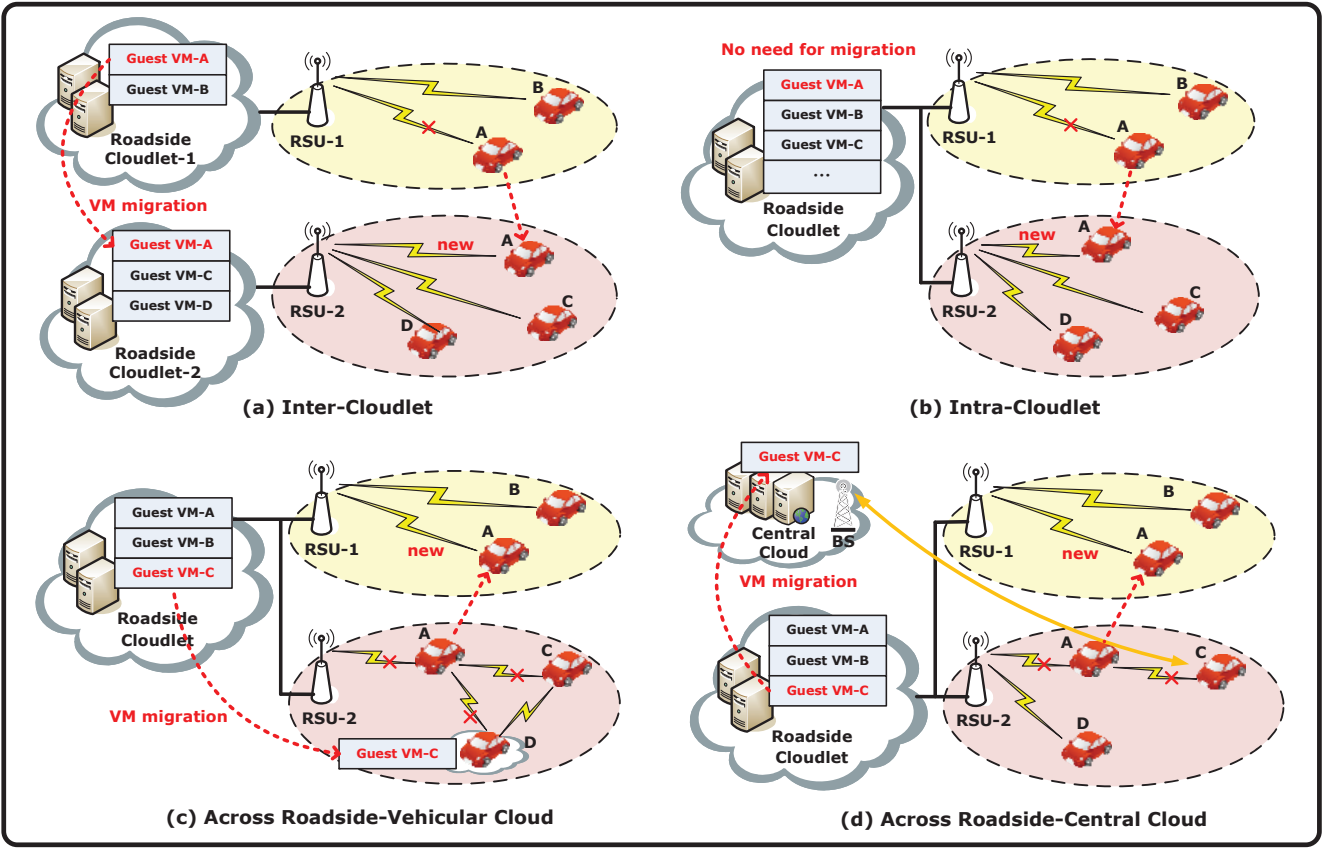}\\
  \caption{Virtual machine migration scenarios}
\end{figure*}

\section{Resource Reservation Scheme for Virtual Machine Migration}

\subsection{Virtual Machine Migration Scenarios}

VM migration refers to the process that an operating VM is transferred along
with its applications across different physical machines. In a VM migration,
a VM image has to be copied from the source to the destination roadside
cloudlets. Different from traditional VM migration, VM migration in
cloud-based vehicular networks has several different scenarios due to
different deployments of roadside clouds and vehicles movements.
\begin{itemize}
  \item \emph{Inter-Cloudlet Case:} In Fig.~4(a), when vehicle $A$ moves from the coverage
  area of RSU-1 to that of RSU-2, a VM migration is needed. Since RSU-1 and RSU-2
  connect to different cloudlets, guest VM-A should be transferred from roadside
  cloudlet-1 to roadside cloudlet-2. After that, $A$ will access cloudlet-2 via
  RSU-2 to resume its service.
  \item \emph{Intra-Cloudlet Case:} In Fig.~4(b), vehicle $A$ moves from the coverage
  area of RSU-1 to that of RSU-2. Since these two RSUs connect to the same roadside
  cloudlet, there is no need for VM migration. However, radio handoff from RSU-1
   to RSU-2 may still take a short period. During handoff, the interaction
   between vehicle $A$ and guest VM-A may be temporally suspended.
  \item \emph{Across Roadside-vehicular cloud Case:} In Fig.~4(c), vehicle $A$ moves from
  the coverage area of RSU-2 to that of RSU-1. Before $A$'s movement, nodes $A$, $C$ and $D$ have
  connection in an ad hoc manner. Vehicle $C$ access the roadside cloud through vehicle $A$.
  The movement of $A$ will cause the disconnection of $C$ from the roadside
  cloud. In this case, guest VM-C will be transferred from the roadside cloud to the vehicle
  cloud in $D$. Then, vehicle $C$ can continue its service through $D$.
    \item \emph{Across Roadside-Central Cloud Case:} The scenario in Fig.~4(d) is similar
    to that in Fig.~4(c), except that there is no direct link between vehicles $C$ and $D$
    in Fig.~4(d). In this case, guest VM-C has to be migrated from the roadside
    cloud to the central
    cloud. After that, $C$ will access the central cloud to resume its service by long-distance
    communications, e.g., 3G/4G cellular.
\end{itemize}

\subsection{Resource Reservation Scheme}

The discussion about VM migration indicates that the VM migration process
involves resource re-allocation in roadside cloud. If the resources of the
destination cloud have been intensively occupied, after a VM migration and
resource re-allocation, some of the VMs may not have sufficient resources and
may not even resume their services. In order to avoid resource
over-commitment, the target cloud site has to deny the VM migration so as to
maintain the services of the existing VMs. In this case, the cloud service of
a vehicle with VM migration is said to be dropped. To reduce service
dropping, we propose a resource reservation scheme. In this scheme, a small
portion of the cloud site resources are reserved merely for migrated VMs, but
not for local VMs. When there are dedicated resources for VM migration, the
dropping rate of cloud services will be significantly decreased.

In the proposed resource reservation scheme, resources are divided into two
categories: reserved resources and common resources. Let $C_r$ and $M_r$
denote the reserved resources, and $C_c=C-C_r$ and $M_c=M-M_r$ the common
resources in computation and storage, respectively. In VM migration, a
\emph{VM arrival} refers to the event that a VM is created either for a new
local VM or a migrated VM. A \emph{VM departure} refers to the request of a
VM deletion, either for an ending of VM service or VM migration to another
cloud site. The resource reservation scheme operates as follows:
\begin{itemize}
  \item \emph{Local VM arrival:} When there is a request for creating a
  new local VM, resource allocation will be carried out, e.g., using the proposed
  game-theoretic allocation scheme. Since a part of the resources are reserved,
  the local VMs can only share the common resources.
  If the resource allocation result satisfies all existing VMs,
   the new local VM is admitted; otherwise, it is blocked.
  \item \emph{Local VM departure:} Resource allocation is also performed
  when the service of a local VM ends or migrates to another cloud site.
  \item \emph{Migrated VM arrival:} Upon a request for a VM migration, the target
  cloud site will re-allocate resources. In this case, the reserved resources will
  be also taken into account. Specifically, the existing local VMs and the migrated VM
  will share all available resources. After re-allocation, if all the VMs
  (including the migrated VM) resource requests are satisfied, VM migration is approved;
  otherwise, the VM migration request is rejected.
  \item \emph{Migrated VM departure:}
      Resource allocation is also performed when the service of a migrated VM ends
      or it migrates to another cloud site. It is noticeable that, if there is no
      migrated VMs in a cloud site, the resource allocation can only use common resources.
      The reserved resources will be conserved for further usage upon another VM migration.
\end{itemize}

\subsection{Optimal Resource Reservation}
We consider $K$ classes of VMs. Let $c_k$ and $m_k$ represent the amount of
required resources by the $k$-th class of VMs in computation and storage,
respectively. Let $n_k^l$ and $n_k^g$ denote the number of local and migrated
VMs of class $k$, respectively. Suppose that the arrivals and departures of
both local and migrated VMs follow a Poisson process model. The system state
transition may be formulated as continuous-time Markov process. Let
$\mathbf{n}_l=(n^l_1,\cdots,n^l_k,\cdots,n^l_K)$ and
$\mathbf{n}_g=(n^g_1,\cdots,n^g_k,\cdots,n^g_K)$. We represent the system state by $\mathbf{s}=(\mathbf{n}_l,\mathbf{n}_g)$ and the state space by $\mathcal{S}$. Let $\pi_\mathbf{s}$ denote the steady state probability of state $\mathbf{s}$. Given the arrival
and departure rates of new and migrated VMs, the steady state probability
matrix $\Pi=\{\pi_\mathbf{s}|\mathbf{s}\in \mathcal{S}\}$ will be derived by a $2K$-dimension Markov chain model.

Let $R_b$ and $R_d$ denote the blocking rate and the dropping rate,
respectively. Then, a new local VM is blocked if the total amount of required resources of the local VMs (including the new one) exceeds that of the common resources, i.e., $\sum_{k=1}^{K}n^l_kc_k > C_c$ or $\sum_{k=1}^{K}n^l_km_k > M_c$. A migrated VM is dropped if the total amount of required resources of all VMs (including the migrated one) is more than that of all resources,
i.e., $\sum_{k=1}^{K}(n^l_k+n^g_k)c_k > C$ or $\sum_{k=1}^{K}(n^l_k+n^g_k)m_k> M$.
Let $\lambda_k^l$, $\mu_k^l$, $\lambda_k^g$ and $\mu_k^g$ denote the arrival and departure rates of local and migrated VMs, then $\mathcal{S}_b$ and $\mathcal{S}_d$ the sets of states that encounter blocking and dropping, respectively. We can derive $R_b(C_r,M_r)=\sum_{\mathbf{s}\in\mathcal{S}_b}\sum_k\pi_\mathbf{s}\lambda_k^l$, and  $R_b(C_r,M_r)=\sum_{\mathbf{s}\in\mathcal{S}_d}\sum_k\pi_\mathbf{s}\lambda_k^g$.
Let $R_b^c$ denote the constraint of the blocking rate. The optimal number of reserved resources is derived by solving the following optimization problem.
\begin{equation}
\begin{split}
\min \quad& R_d(C_r,M_r),\\
\mathrm{s.t.}\quad&R_b(C_r,M_r)\leq R_b^c.
\end{split}
\end{equation}

\begin{figure}[t]
\centering
  \includegraphics[width=0.5\textwidth]{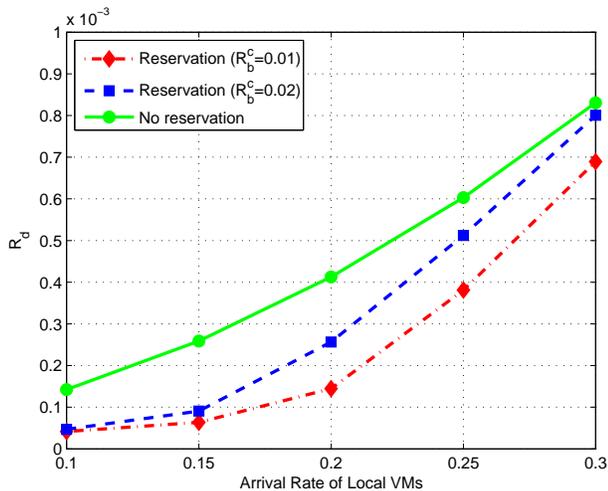}\\
  \caption{Dropping rate in terms of local VM arrival rate}
\end{figure}

Fig.5 shows a performance comparison with and without resource reservation.
The total resources of the roadside clouds are $50$ and $100$ units in
computation and storage, respectively. Two classes of VMs are considered. VMs of class-1 are mainly for computation-type applications, which needs $20$ units in computational resources and $15$ units in storage resources. VMs of class-2 are mainly for storage-type applications, which need $10$ units in computational resources and $40$ units in storage resources. 
The two classes of VMs are assumed to have identical in arrival and departure rates. We set the range of local VM arrival rate from $0.1$ to $0.3$, the local VM departure rate by $2.0$, the arrival and departure rates of migrated VM by $0.05$ and $0.1$, respectively.
The simulation results show that the dropping rate of migrated VMs is significantly reduced with resource reservation, which demonstrates the efficiency of our proposed mechanism.

\section{Conclusions}

In this article, we first discussed the opportunities and challenges in
exploiting cloud computing in vehicular networks. Then, we presented a
hierarchical architecture for cloud-based vehicular networks that facilitates sharing of computational resources, storage resources and bandwidth resources among vehicles.
Furthermore, we focused on efficient resource management in the proposed
architecture. The resource competition among virtual machines is formulated
and solved in a game-theoretical framework. Virtual resource migration due to
vehicle mobility is addressed based on a resource reservation scheme.
Finally, illustrative results indicated a significant reduction of the
service dropping rate during virtual machine migration.

\section*{Acknowledagement}

This research is partially supported by program of NSFC (grant
no.~U1035001, U1201253, 61203117), the Opening Project
of Key Lab. of Cognitive Radio and Information Processing
(GUET), Ministry of Education (grant no. 2011KF06),
the project 217006 funded by the Research Council of Norway, the European Commission FP7 Project EVANS (grant no. 2010-269323), and the European Commission COST Action IC0902, IC0905 and IC1004.

\begin{IEEEbiographynophoto}{Rong Yu}
[S¡¯05, M¡¯08] (yurong@ieee.org) received his Ph.D. from Tsinghua University, China, in 2007. After that, he worked in the School of Electronic and Information Engineering of South China University of Technology (SCUT). In 2010, he joined the Institute of Intelligent Information Processing at Guangdong University of Technology (GDUT), where he is now an associate professor. His research interest mainly focuses on wireless communications and networking, including cognitive radio, wireless sensor networks, and home networking. He is the co-inventor of over 10 patents and author or co-author of over 50 international journal and conference papers.
Dr. Yu is currently serving as the deputy secretary general of the Internet of Things (IoT) Industry Alliance, Guangdong, China, and the deputy head of the IoT Engineering Center, Guangdong, China. He is the member of home networking standard committee in China, where he leads the standardization work of three standards.
\end{IEEEbiographynophoto}

\begin{IEEEbiographynophoto}{Yan Zhang}
[SM¡¯10] (yanzhang@ieee.org) received a Ph.D. degree from
Nanyang Technological University, Singapore. He is working with Simula
Research Laboratory, Norway; and he is an adjunct Associate Professor at the
University of Oslo, Norway. He is an associate editor or guest editor of a number of international journals. He serves as organizing committee chairs for many international conferences. His research interests include resource, mobility, spectrum, energy, and data management in wireless communications and networking.
\end{IEEEbiographynophoto}

\begin{IEEEbiographynophoto}{Stein Gjessing}
(steing@ifi.uio.no) is a professor of computer science in Department
of Informatics, University of Oslo and an adjunct researcher at Simula
Research Laboratory. He received his Dr. Philos. degree from the University of Oslo in 1985. Gjessing acted as head of the Department of Informatics for 4 years from 1987. From February 1996 to October 2001 he was the chairman
of the national research program ¡°Distributed IT-System,¡± founded by the
Research Council of Norway. Gjessing participated in three European funded
projects: Macrame, Arches and Ascissa. His current research interests are routing, transport protocols and wireless networks, including cognitive radio and smart grid applications.
\end{IEEEbiographynophoto}

\begin{IEEEbiographynophoto}{Wenlong Xia}
(wenlong.xia@ieee.org) received his M.S. degree in electronics engineering from the PLA Information Engineering University, China in 2011. Now he is pursuing his M.S. degree in signal and information processing from Guangdong University of Technology (GDUT), China. His research interests include vehicular wireless networks, opportunistic networks and cloud computing.
\end{IEEEbiographynophoto}

\begin{IEEEbiographynophoto}{Kun Yang}
(kunyang@essex.ac.uk) received his PhD from the Department of Electronic \& Electrical Engineering of University College London (UCL), UK. He is currently a full Professor in the School of Computer Science \& Electronic Engineering, University of Essex, UK, and the Head of the Network Convergence Laboratory (NCL) in Essex. His main research interests include wireless networks/communications, fixed mobile convergence, future Internet technology and network virtualization. He has published over 150 papers in the above research areas. He serves on the editorial boards of both IEEE and non-IEEE journals. He is a Senior Member of IEEE and a Fellow of IET.
\end{IEEEbiographynophoto}

\end{document}